\documentclass[aps,preprint,prb,showpacs]{revtex4-1}%
\usepackage[colorlinks=true,linkcolor=blue]{hyperref}
\usepackage{amssymb}
\usepackage{amsfonts}
\usepackage{amsmath}
\usepackage{graphicx}%
\providecommand{\U}[1]{\protect\rule{.1in}{.1in}}

\begin{document}
\preprint{REV\TeX4-1 }
\title[Geometric Momentum on Curved Surface]{Geometric Momentum for a Particle on a Curved Surface}
\author{Q. H. Liu}
\email{quanhuiliu@gmail.com}
\affiliation{School for Theoretical Physics, and Department of Applied Physics, Hunan
University, Changsha, 410082, China}

\begin{abstract}
When a two-dimensional curved surface is conceived as a limiting case of a
curved shell of equal thickness $d$, where the limit $d\rightarrow0$ is then
taken, the well-known \emph{geometric potential} is induced by the kinetic
energy operator, in fact by the second order partial derivatives. Applying
this confining procedure to the momentum operator, in fact to the first order
partial derivatives, we find the so-called \emph{geometric momentum }instead.
This momentum is compatible with the Dirac's canonical quantization theory on
system with second-class constraints. The distribution amplitudes of the
geometric momentum on the spherical harmonics are analytically determined, and
they are experimentally testable for rotational states of spherical molecules
such as $C_{60}$.

\end{abstract}
\date{\today}

\maketitle

\textsc{Introduction }The affirmative experimental evidence in 2010
\cite{Szameit} of the \emph{geometric potential} firstly explored in 1971
\cite{jk} and fundamentally finished in 1981 \cite{dacosta} and with correct
inclusion of electromagnetic field in 2008 \cite{FC} etc. \cite{BP} is a
groundbreaking advance of quantum mechanics applied for curved nanostructures,
starting from the three dimensional (3D) bulk system and then reducing it to a
2D surface one. \cite{jk,dacosta,FC,BP} This success echoes a historical
footnote in Dirac's $\mathit{Principle}$ on the canonical quantization
assumption that "is found in practice successful only when applied with the
dynamic coordinates and momenta referring to a Cartesian system of axes and
not to more general curvilinear coordinates." \cite{dirac1} However, the
classic work by Jensen, Koppe \cite{jk} and da Costa \cite{dacosta} hides an
important physical and mathematical message when dealing with derivatives on a
2D curved surface $S$:\ There is a noninterchangeability of order of taking
two limits. Explicitly, when the 2D curved surface is conceived as a limiting
case of a curved shell of equal thickness $d$, where the limit $d\rightarrow0$
is then taken, great discrepancies present as firstly taking limit
$d\rightarrow0$ then defining the derivatives on the surface, and as firstly
defining derivatives in bulk then letting $d\rightarrow0$. The second order is
named as the \emph{confining procedure} for studying motion on 2D surface
embedded in 3D. \cite{FC} For the former order, the quantum kinetic energy
operator is hypothesized to be proportional to Laplace-Beltrami operator
$\Delta_{LB}$ on the surface: \cite{dewitt}%
\begin{equation}
T=-\frac{\hbar^{2}}{2\mu}\Delta_{LB}, \label{1T}%
\end{equation}
whereas for the latter order, we have following form of the kinetic energy
operator $T$ instead: \cite{jk,dacosta,FC}
\begin{equation}
T=-\frac{\hbar^{2}}{2\mu}\Delta_{LB}-\frac{\hbar^{2}}{2\mu}(M^{2}-K),
\label{2T}%
\end{equation}
where $M$ is the mean curvature and $K$ is the gaussian curvature, and the
excess term is called \emph{geometric potential }$V_{gp}$, \cite{Szameit,FC}%
\emph{ }%
\begin{equation}
V_{gp}=-\frac{\hbar^{2}}{2\mu}(M^{2}-K). \label{GP}%
\end{equation}
The experimental verification of this potential implies that the original
Laplace-Beltrami operator $\Delta_{LB}$ on the 2D surface may not be enough
unless a term $(M^{2}-K)$ is included, \cite{Szameit,Schultheiss}
\begin{equation}
\Delta_{LB}\rightarrow\Delta_{LB}+(M^{2}-K). \label{GP1}%
\end{equation}
For avoiding confusion, we adhere to the convention in mathematics where the
Laplace operator $\nabla^{2}$ acting on a function is defined by the
divergence of the gradient of the function in flat space: $\nabla^{2}%
\equiv\nabla\cdot\nabla$, while the Laplace-Beltrami operator $\Delta_{LB}$ is
a generalization of the Laplace operator on surface under consideration.

The noninterchangeability of calculus order must be fundamentally associated
with the gradient $\nabla$, or the momentum operator $\mathbf{p}=-i\hbar
\nabla$, which has not explored before. On the other hand, we must mention an
entirely independent development on the quantization of the momentum on 2D
surface embedded in 3D flat space, \cite{liu07} and the momentum is found to
assume the following form,
\begin{equation}
\mathbf{p}=-i\hbar(\mathbf{r}^{\mu}\partial_{\mu}+M\mathbf{n}), \label{GM}%
\end{equation}
where we use the tensor covariant and contravariant components and the
Einstein summation convention, and
\begin{equation}
\mathbf{r}(q^{1},q^{2})\mathbf{=(}\text{ }x(q^{1},q^{2}),y(q^{1}%
,q^{2}),z(q^{1},q^{2})\text{ }\mathbf{)} \label{standform}%
\end{equation}
is the position vector on the surface $S$ parametrized by ($q^{1},q^{2}$)
denoted by $q^{\mu}$ and $q^{\nu}$ with lowercase greek letters $\mu,\nu$
taking values $1,2$, and $\mathbf{r}^{\mu}=g^{\mu\nu}\mathbf{r}_{\nu}%
=g^{\mu\nu}{\partial_{\nu}}\mathbf{r}$ $=g^{\mu\nu}\partial\mathbf{r/}q^{\nu}$
with $g_{\mu\nu}={\partial_{\mu}}\mathbf{r}\cdot{\partial_{\nu}}\mathbf{r}$
being the metric tensor. At this point $\mathbf{r}$, $\mathbf{n=(}n_{x}%
,n_{y},n_{z}\mathbf{)}$ is the normal and $M\mathbf{n}$ symbolizes the mean
curvature vector field, a geometric invariant. \cite{liu07}

The first aim of the present study is to show that the application of the same
confining procedure pioneered by Jensen, Koppe \cite{jk} and da Costa
\cite{dacosta} to operator $\mathbf{p}=-i\hbar\nabla$ that holds true in bulk
automatically results in $\mathbf{p}=-i\hbar(\mathbf{r}^{\mu}\partial_{\mu
}+M\mathbf{n})$ (\ref{GM}) on the surface. So, analogue to the name
\emph{geometric potential} we can call $\mathbf{p}=-i\hbar(\mathbf{r}^{\mu
}\partial_{\mu}+M\mathbf{n})$ (\ref{GM}) \emph{geometric momentum} (GM).
Because $\mathbf{r}(q^{1},q^{2})$ (\ref{standform}) in mathematics offers the
so-called standard parametrization of the 2D surface, the corresponding GM
(\ref{GM}) should offer proper description of the momentum. If simply denoting
the gradient operator $\mathbf{r}^{\mu}\partial_{\mu}$\ \cite{oy} on the
surface by $\nabla_{//}$, Eq. (\ref{GM}) implies following correspondence,
\begin{equation}
\nabla_{//}\rightarrow\nabla_{//}+M\mathbf{n.} \label{GM1}%
\end{equation}
On a surface, is there a component $M\mathbf{n}$ normal to it? This result
(\ref{GM1}) is somewhat contrary to what physical intuition or common sense
would indicate. But it is the case as examined in 3D flat space.

\textsc{geometric momentum as a consequence of confining procedure } To prove
the GM (\ref{GM}), we utilize exactly the same manner how to derive the
geometric potential. \cite{jk,dacosta,FC} For ease of the comparison, we use
similar set of symbols as Ferrari and Cuoghi who recently build up a
theoretical framework with geometric potential when the electromagnetic field
is present. \cite{FC} The lowercase Latin letters $i,j,k$ stand for the 3D
indices and assume the values $1,2,3$, e.g., $\mathbf{(}x_{i},p_{j}\mathbf{)}$
for the position and momentum. Position specified by ($q^{1},q^{2},q^{3}$) can
be understood as description of the position in the curvilinear coordinates
parameterizing a manifold. The original 2D surface $\mathbf{r}(q^{1},q^{2})$
is considered as a more realistic 3D shell whose equal thickness $d$ is
negligible in comparison with the dimension of the whole system. The position
$\mathbf{R}$ within the shell in the vicinity of the surface $S$ can be
parametrized as with $0$ $\leq q^{3}\leq d,$
\begin{equation}
\mathbf{R}(q^{1},q^{2},q^{3})=\mathbf{r}(q^{1},q^{2})+q^{3}\mathbf{n}%
(q^{1},q^{2}).\label{3dR}%
\end{equation}
The gradient operator $\nabla$ in 3D flat space, expressed in the curvilinear
coordinates, takes following form, \cite{oy}
\begin{equation}
\nabla=\mathbf{r}^{\mu}\partial_{\mu}+\mathbf{n}\partial_{q^{3}}%
\text{.}\label{grad}%
\end{equation}
The relation between the 3D metric tensor $G_{ij}$ and the 2D one $g_{\mu\nu}$
is given by, \cite{dacosta,FC}
\begin{align}
G_{ij} &  =g_{\mu\nu}+\left[  \alpha g+(\alpha g)^{T}\right]  _{\mu\nu}%
q^{3}+(\alpha g\alpha^{T})_{\mu\nu}\left(  q^{3}\right)  ^{2},\nonumber\\
G_{\mu3} &  =G_{3\mu}=0,\;G_{33}=1,\label{metric}%
\end{align}
where $\alpha_{\mu\nu}$ is the Weingarten curvature matrix for the surface,
and $M=-\mathrm{Tr}(\alpha)/2$, and $K=\mathrm{\det}(\alpha)$. \cite{dacosta}
The covariant Schr\"{o}dinger\ equation for particles moving within a thin
shell of thickness $d$ in 3D is, with both the vector potential $V$ and the
electric potential $\mathbf{A}$ applied, \cite{FC}
\begin{equation}
\mathrm{i}\hbar\frac{\partial}{\partial t}\psi(\mathbf{q},t)=-\frac{\hbar^{2}%
}{2m}G^{ij}D_{i}D_{j}\psi(\mathbf{q},t)+QV\psi(\mathbf{q},t),\label{schr1}%
\end{equation}
where $Q$ is the charge of the particle and $D_{j}=\nabla_{j}-(\mathrm{i}%
Q/\hbar)A_{j}$ with $A_{j}$ being the covariant components of the vector
potential $\mathbf{A}$. Defining the scalar potential $A_{0}=-V$, we can
define a gauge covariant derivative for the time variable as $D_{0}%
={\partial_{t}}-{\mathrm{i}Q}A_{0}/\hbar$, and rewrite Eq.(\ref{schr1}) as,
\cite{FC}
\begin{equation}
\mathrm{i}\hbar D_{0}\psi=-\frac{\hbar^{2}}{2m}G^{ij}D_{i}D_{j}\psi
.\label{schr2}%
\end{equation}
This equation is evidently gauge invariant with respect of the following gauge
transformations: \cite{FC}
\begin{equation}
A_{j}\rightarrow A_{j}^{\prime}=A_{j}+\partial_{j}\gamma;\text{ }%
A_{0}\rightarrow A_{0}^{\prime}=A_{0}+{\partial_{t}}\gamma;\text{ }%
\psi\rightarrow\psi^{\prime}=\psi\mathrm{e}^{\mathrm{i}Q\gamma/\hbar
},\label{GTran}%
\end{equation}
where $\gamma$ is a scalar function.

Two important facts regarding the wave functions will be needed. 1, the
normalization of the wave functions remains whatever coordinates are used, and
we have with transformation of volume element $d^{3}\mathbf{x=}\sqrt{G}%
d^{3}\mathbf{q}$, \cite{FC}
\begin{equation}
\int\left\vert \psi(\mathbf{x},t)\right\vert ^{2}\mathrm{d}^{3}\mathbf{x=}%
\int\left\vert \psi(\mathbf{q},t)\right\vert ^{2}\sqrt{G}\mathrm{d}%
^{3}\mathbf{q}=1,\label{Totalnormal}%
\end{equation}
where \cite{dacosta,FC}
\begin{equation}
G=\mathrm{det}(G_{ij})=g\left(  1-2Mq^{3}+K\left(  q^{3}\right)  ^{2}\right)
^{2}.
\end{equation}
2, an advantage of the curvilinear coordinates is the accessibility of the
separability of the wave function $\psi(\mathbf{q},t)$ in (\ref{schr1}) or
(\ref{schr2}) as, \cite{dacosta,FC}
\begin{equation}
\psi(\mathbf{q},t)=\frac{\chi(q^{1},q^{2},t)}{\sqrt{1-2Mq^{3}+K\left(
q^{3}\right)  ^{2}}}\varphi(q^{3},t),
\end{equation}
and it is guaranteed with suitable choice of gauge for $\gamma$ such that
$A_{3}^{\prime}=0$, \cite{FC}
\begin{equation}
\gamma(q^{1},q^{2},q^{3})=-\int_{0}^{q^{3}}A_{3}(q^{1},q^{2},q)\mathrm{d}%
q.\label{ggauge}%
\end{equation}
Combining these two facts, we have two conservations of norm from
(\ref{Totalnormal}),%
\begin{gather}
\oint\left\vert \chi(q^{1},q^{2},t)\right\vert ^{2}\sqrt{g}\mathrm{d}%
q^{1}\mathrm{d}q^{2}=1,\\
\text{and }\int_{0}^{d}\left\vert \varphi(q^{3},t)\right\vert ^{2}%
\mathrm{d}q^{3}=1.
\end{gather}

We are now ready to examine the gradient operator $\nabla$ (\ref{grad}) acting
on the $\psi(\mathbf{q},t)$ and the result is,%
\begin{align}
\nabla\psi(\mathbf{q},t) &  =\mathbf{r}^{\mu}\partial_{\mu}\psi(\mathbf{q}%
,t)+\mathbf{n}\frac{M+q^{3}K}{\left(  1-2Mq^{3}+K\left(  q^{3}\right)
^{2}\right)  ^{3/2}}\chi(q^{1},q^{2},t)\varphi(q^{3},t)\nonumber\\
&  +\mathbf{n}\frac{\chi(q^{1},q^{2},t)}{\sqrt{1-2Mq^{3}+K\left(
q^{3}\right)  ^{2}}}\partial_{q^{3}}\varphi(q^{3},t).
\end{align}
Then taking limit $d\rightarrow0$, we have,
\begin{equation}
\nabla\psi(\mathbf{q},t)=\left(  \mathbf{r}^{\mu}\partial_{\mu}+M\mathbf{n}%
\right)  \psi(\mathbf{q},t)+\mathbf{n}\chi(q^{1},q^{2},t)\partial_{q^{3}%
}\varphi(q^{3},t),\label{dd}%
\end{equation}
which implies that the gradient operator $\nabla$ can be decomposed into two
separate parts, one is ($q^{1},q^{2}$) dependent part $\left(  \mathbf{r}%
^{\mu}\partial_{\mu}+M\mathbf{n}\right)  $ and another the $q^{3}$-derivative
part $\mathbf{n}\partial_{q^{3}}$, corresponding to the decomposition of the
Schr\"{o}dinger equation into two Schr\"{o}dinger ones determining $\chi
(q^{1},q^{2},t)$ and $\varphi(q^{3},t)$ respectively. \cite{FC} Paying
attention to the motion on the surface only, we have the resultant operator
$\mathbf{r}^{\mu}\partial_{\mu}+M\mathbf{n}$ (\ref{GM1}). In fact, with proper
choice of the confining potential $V(q^{3})$ in the confining procedure,
\cite{dacosta,FC} we can require that $\int_{0}^{d}\varphi^{\ast}%
(q^{3},t)\partial_{q^{3}}\varphi(q^{3},t)\mathrm{d}q^{3}=0$. So, after
performing an integration of operator $\nabla$ in (\ref{dd}) over
perpendicular interval $[0,d]$ as $\int_{0}^{d}\varphi^{\ast}(q^{3}%
,t)\nabla\varphi(q^{3},t)\mathrm{d}q^{3}$, only the surface part $\left(
\mathbf{r}^{\mu}\partial_{\mu}+M\mathbf{n}\right)  $ (\ref{GM1}) survives.

The gauge invariance of the momentum operator $\mathbf{p}=-i\hbar
(\mathbf{r}^{\mu}\partial_{\mu}+M\mathbf{n)-}Q\mathbf{A}$ is assured in the
presence of the vector potential $\mathbf{A}$ with 3D gauge $A_{3}=0$ being
pre-imposed. Under 2D gauge transformation: $\mathbf{A}\rightarrow
\mathbf{A}^{\prime}=\mathbf{A+r}^{\mu}\partial_{\mu}\gamma$ with
$\gamma=\gamma(q^{1},q^{2})$ and $\psi\rightarrow\psi^{\prime}=\mathrm{e}%
^{\mathrm{i}Q\gamma/\hbar}\psi$, we have $\mathbf{p}\psi\rightarrow
\mathbf{p}^{\prime}\psi^{\prime}=\mathrm{e}^{\mathrm{i}Q\gamma/\hbar
}\mathbf{p}\psi$,
\begin{align}
\mathbf{p}^{\prime}\psi^{\prime}  &  =\left(  -i\hbar(\mathbf{r}^{\mu}%
\partial_{\mu}+M\mathbf{n)-}Q(\mathbf{A}+\mathbf{r}^{\mu}\partial_{\mu}%
\gamma)\right)  \psi\mathrm{e}^{\mathrm{i}Q\gamma/\hbar}\nonumber\\
&  =\mathrm{e}^{\mathrm{i}Q\gamma/\hbar}\left(  -i\hbar(\mathbf{r}^{\mu
}\partial_{\mu}+M\mathbf{n)-}Q\mathbf{A}\right)  \psi\nonumber\\
&  =\mathrm{e}^{\mathrm{i}Q\gamma/\hbar}\mathbf{p}\psi.
\end{align}

So far, we also understand why\ there is no direct connection between
$\Delta_{LB}+(M^{2}-K)$ and $\nabla_{//}+M\mathbf{n}$ such as in 3D flat space
$\nabla^{2}\equiv\nabla\cdot\nabla$. For reaching $\Delta_{LB}+(M^{2}-K)$, we
have to start from the Laplace operator in flat 3D space$\ \nabla^{2}=\left(
\mathbf{r}^{\mu}\partial_{\mu}+\mathbf{n}\partial_{q^{3}}\right)  \cdot\left(
\mathbf{r}^{\mu}\partial_{\mu}+\mathbf{n}\partial_{q^{3}}\right)  =\Delta
_{LB}+M\partial_{q^{3}}+\partial_{q^{3}}^{2}$, then resort to the confining procedure.

\textsc{geometric momentum in Dirac's theory }During 1950's and 1960's, Dirac
\cite{dirac2} establishes a theory for constrained motion instead follows the
routine paradigm of quantization hypothesis of kinetic energy $T=-\hbar
^{2}(2\mu)\Delta_{LB}$ on the curved surface. Recalling his famous footnote,
\cite{dirac1} we can reasonably infer that if his understanding of canonical
quantization is self-consistent and indeed insightful, the geometric momentum
(\ref{GM}) must be a natural realization of the momentum in the Dirac's
canonical quantization for a system with second-class constraints. The second
aim of the present study is to illustrate that it is really the case.

For the constrained motion on the surface $S$ (\ref{standform}), Dirac's
theory gives for the commutators: \cite{homma,ikegami,okamoto,golovnev}%
\begin{equation}
\lbrack x_{i},p_{j}]=\mathrm{i}\hbar(\delta_{ij}-n_{i}n_{j}),\text{
}[\mathbf{r},T]=\mathrm{i}\hbar\frac{\mathbf{p}}{m}. \label{dirac}%
\end{equation}
The verification of the first commutator whose tensor form is $[\mathbf{r}%
,\mathbf{p}]=\mathrm{i}\hbar(\overset{\rightarrow\rightarrow}{\mathbf{I}%
}-\mathbf{nn})$ needs an identity for\ the second-rank tensor\ as
$\overset{\rightarrow\rightarrow}{\mathbf{I}}\mathbf{=nn+r}^{\mu}%
\mathbf{r}_{\mu}$ whose proof is straightforward. In fact, the tensor form of
the commutator $[\mathbf{r},\mathbf{p}]\equiv\mathbf{rp-pr}$ gives:
\begin{equation}
\lbrack\mathbf{r},\mathbf{p}]\equiv-\mathrm{i}\hbar\lbrack\mathbf{r,}%
(\mathbf{r}^{\mu}\partial_{\mu}+M\mathbf{n)}]=\mathrm{i}\hbar\mathbf{r}^{\mu
}\mathbf{r}_{\mu}=\mathrm{i}\hbar(\overset{\rightarrow\rightarrow}{\mathbf{I}%
}-\mathbf{nn}).
\end{equation}
The second commutator is evident with use of a formula $\nabla^{2}%
\mathbf{r}=2M\mathbf{n}$, \cite{oy}%
\begin{align}
\lbrack\mathbf{r},T]  &  =-\frac{\hbar^{2}}{2m}[\mathbf{r,}\Delta_{LB}%
]=\frac{\hbar^{2}}{2m}\left(  \frac{1}{\sqrt{g}}(\partial_{\mu}g^{\mu\upsilon
}\sqrt{g}\partial_{\upsilon}\mathbf{r)+}2\mathbf{(}\partial^{\upsilon
}\mathbf{r)}\partial_{\upsilon}\right) \nonumber\\
&  =\frac{\hbar^{2}}{2m}\left(  2M\mathbf{n+}2\mathbf{r}^{\mu}\partial_{\mu
}\right) \nonumber\\
&  =\frac{i\hbar}{m}\mathbf{p.}%
\end{align}

\textsc{geometric momentum distribution} \textsc{of spherical harmonics} The
third aim of the present study is to give the probability distribution of
geometric momentum of the spherical harmonics $Y_{lm}(\theta,\varphi)$. For
our propose, we firstly give GM for particle on the surface of unit sphere.
\cite{liu07,liu03,liu10}
\begin{align}
p_{x}  &  =-i\hbar(\cos\theta\cos\varphi\frac{\partial}{\partial\theta}%
-\frac{\sin\varphi}{\sin\theta}\frac{\partial}{\partial\varphi}-\sin\theta
\cos\varphi),\label{hpx}\\
p_{y}  &  =-i\hbar(\cos\theta\sin\varphi\frac{\partial}{\partial\theta}%
+\frac{\cos\varphi}{\sin\theta}\frac{\partial}{\partial\varphi}-\sin\theta
\sin\varphi),\label{hpy}\\
p_{z}  &  =i\hbar(\sin\theta\frac{\partial}{\partial\theta}+\cos\theta).
\label{hpz}%
\end{align}
These operators satisfy the definition of the vector operator \cite{Sakurai}
as $[{L}_{{i}}{,p}_{j}]=i\hbar\varepsilon_{ijk}{p}_{k}$, we can therefore have
the operators $p_{x}$ and $p_{y}$ from $p_{z}$ by means of rotation of the
axis' rotation. Explicitly, rotation $\pi/2$ around $y$-axis renders $p_{z}$
to be $p_{x}$, and $-\pi/2$ around $x$-axis renders $p_{z}$ to be $p_{y}$,
\begin{equation}
p_{x}=\exp(-i\pi L_{y}/2)p_{z}\exp(i\pi L_{y}/2),\text{ }p_{y}=\exp(i\pi
L_{x}/2)p_{z}\exp(-i\pi L_{x}/2). \label{rotation}%
\end{equation}
Here we follow the convention that a rotation operation affect a physical
system itself. \cite{Sakurai} Hence the eigenvalue problem for operators
$p_{x}$ or $p_{y}$ is simultaneously determined once the complete solution to
$\hat{p}_{z}\psi_{p_{z}}(\theta)=p_{z}\psi_{p_{z}}(\theta)$ is known, where on
operator $p_{z}$ on the left hand side of this equation the carat symbol
"$\symbol{94}$" is used to distinguish the eigenvalue $p_{z}$ on the right
hand side. The eigenfunctions $\psi_{p_{z}}(\theta)$ form a complete set once
the eigenvalues $p_{z}$ are real and continuous,
\begin{equation}
\psi_{p_{z}}(\theta)=\frac{1}{2\pi}\frac{1}{\sin\theta}\tan^{-ip_{z}}\left(
\frac{\theta}{2}\right)  . \label{eigenf}%
\end{equation}
They are $\delta$-function normalized,
\begin{align}
&  \oint{\psi_{{{{{p}^{\prime}}}_{z}}}^{\ast}}\left(  \theta,\varphi\right)
{{\psi}_{{{p}_{z}}}}\left(  \theta,\varphi\right)  \sin\theta d\theta
d\varphi\nonumber\\
&  =\frac{1}{2\pi}\int_{0}^{\pi}{\exp\left(  i\left(  {{{{p}^{\prime}}}_{z}%
}-{{p}_{z}}\right)  (\ln\tan\frac{\theta}{2})\right)  }\frac{1}{\sin\theta
}d\theta\nonumber\\
&  =\frac{1}{2\pi}\int_{0}^{\pi}{\exp\left(  i\left(  {{{{p}^{\prime}}}_{z}%
}-{{p}_{z}}\right)  \ln\tan\frac{\theta}{2}\right)  }d\ln\tan\frac{\theta}%
{2}\nonumber\\
&  =\frac{1}{2\pi}\int_{-\infty}^{\infty}{\exp\left(  i\left(  {{{{p}^{\prime
}}}_{z}}-{{p}_{z}}\right)  z\right)  }dz\nonumber\\
&  =\delta\left(  {{{{p}^{\prime}}}_{z}}-{p}_{{z}}\right)  ,
\label{normalization}%
\end{align}
where the variable transformation $\ln\tan\theta/2\rightarrow z$ is used. So,
we see explicitly that the eigenfunctions $\psi_{p_{z}}(\theta)$ form a
complete set. Next we use it to expand the spherical harmonics $Y_{lm}%
(\theta,\varphi)$. Because of the symmetry, the momentum distribution along
$z$-axis depends on the angular quantum number $l$ only. The result turns out
to be,%
\begin{align}
\varphi_{l}(p_{z})  &  =%
{\displaystyle\oint}
Y_{lm}(\theta,\varphi){\psi_{{{{{p}}}_{z}}}^{\ast}}\left(  \theta
,\varphi\right)  \sin\theta\mathrm{d}\theta\mathrm{d}\varphi\nonumber\\
&  =\sqrt{\frac{2l+1}{2}}F(p_{z})\left[  \frac{P_{l}(\tanh q)}{\cosh
q}\right]  ,
\end{align}
where $P_{l}$ is the Legendre function of order $l$ and the Fourier transform
$F(p)\left[  f(q)\right]  $ of a function $f(q)$ is defined by,%
\begin{equation}
F(p)\left[  f(q)\right]  \equiv%
{\displaystyle\int}
f(q)\frac{\mathrm{e}^{-\mathrm{i}pq}}{\sqrt{2\pi}}\mathrm{d}q.
\end{equation}

The first three momentum distribution $\varphi_{l}(p_{z})$ are respectively,%
\begin{align}
\varphi_{0}(p_{z})  &  =\frac{\sqrt{\pi}}{2}\sec\mathrm{h}\left(  \frac{\pi
p_{z}}{2}\right)  ,\\
\varphi_{1}(p_{z})  &  =\frac{\mathrm{i}\sqrt{3\pi}}{2}p_{z}\sec
\mathrm{h}\left(  \frac{\pi p_{z}}{2}\right)  ,\\
\varphi_{2}(p_{z})  &  =\frac{\sqrt{5\pi}}{8}(3p_{z}^{2}-1)\sec\mathrm{h}%
\left(  \frac{\pi p_{z}}{2}\right)  .
\end{align}
The ground state $Y_{00}(\theta,\varphi)$ is the minimum uncertainty state for
three pairs of ($x_{i},p_{i}$) and $\Delta x_{i}\Delta p_{i}=\hbar/3$. In zero
angular momentum state $Y_{00}(\theta,\varphi)$ that bears no energy either,
the presence of zero-point the momentum fluctuation $\overline{\left(  \Delta
p_{i}\right)  ^{2}}=\hbar^{2}/3$ contradicts what classical mechanics would
indicate. In overall respects, these states bears striking resemblance to the
probability amplitude of the momentum for one-dimensional simple harmonic
oscillator. The momentum distributions in the spherical harmonics offers an
experimentally testable result for rotational state of spherical molecule such
as $C_{60}$. With preparing these molecules into ground state of rotation, the
probability of the momentum distributions is depicted in Fig. 1.

\begin{figure}
\includegraphics[scale=0.9]{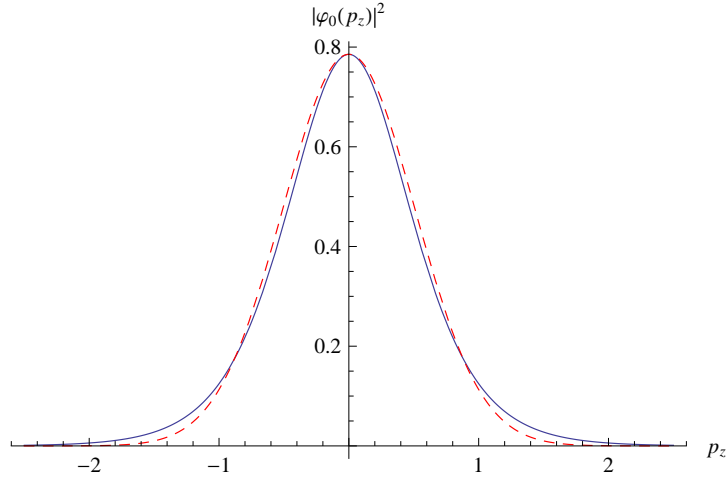}\caption{Momentum distribution density for $Y_{00}(\theta,\varphi)$ (solid line) and for the ground state of 1D simple harmonic
oscillator (dashed line) are plotted. They are almost identical.}\label{fig1}
\end{figure}

\textsc{Conclusions and discussions }Bye use of the same confining procedure
pioneered by Jensen, Koppe \cite{jk} and da Costa \cite{dacosta} to give the
geometric potential, we find that the momentum $\mathbf{p}=-i\hbar\nabla$
originally defined in bulk becomes a momentum $\mathbf{p}=-i\hbar
(\mathbf{r}^{\mu}\partial_{\mu}+M\mathbf{n})$ (\ref{GM}) defined on the
surface, which was previously proposed on completely different ground.
Remarkably, this momentum is compatible with the Dirac's canonical
quantization theory on system with second-class constraints. Because
$\mathbf{r}(q^{1},q^{2})$ (\ref{standform}) in mathematics offers the
so-called standard parametrization of the 2D surface, the corresponding
momentum $\mathbf{p}$ (\ref{GM}) should be also preferable over other forms of
momentum such as the generalized momenta ($p_{q^{1}},p_{q^{2}}$) canonically
conjugated to parameters ($q^{1},q^{2}$). This is another reason it deserves a
terminology, \emph{geometric momentum} as we called. The distribution
amplitudes of the geometric momentum of the spherical harmonics are
analytically determinable, and experimentally testable for rotational state of
spherical molecule such as $C_{60}$.

\begin{acknowledgments}
This work is financially supported by National Natural Science Foundation of
China under Grant No. 11175063, and by Program for New Century Excellent
Talents in University, Ministry of Education, China.
\end{acknowledgments}

\end{document}